\documentclass{emulateapj}
\usepackage{apjfonts}

\newcommand {\bband} {B_{\rm F435W} } 
\newcommand {\vband} {V_{\rm F606W} } 
\newcommand {\iband} {I_{\rm F775W} } 
\newcommand {\zband} {Z_{\rm F850LP} } 
\newcommand {\solarmassperyear}  {{\rm M_{\sun}~yr^{-1}}}

\shorttitle{Extinction in Lyman-Break Galaxies}
\shortauthors{To et al.}

\begin{document}
\title{Star Formation Rate and Extinction in Faint $z\sim4$ Lyman-Break Galaxies}

\author{Chun-Hao To\altaffilmark{1,2}, Wei-Hao Wang\altaffilmark{1}, and Frazer N.\ Owen\altaffilmark{3}}

\altaffiltext{1}{Academia Sinica Institute of Astronomy and Astrophysics, P.O. Box 23-141, Taipei 10617, Taiwan}
\altaffiltext{2}{Department of Physics, National Taiwan University, No.\ 1, Sec.\ 4, Roosevelt Rd., Taipei 106, Taiwan}
\altaffiltext{3}{National Radio Astronomy Observatory, P.O. Box 0, Socorro, NM 87801, USA}

\begin{abstract}
We present a statistical detection of 1.5 GHz radio continuum emission from a sample of faint $z\sim4$ Lyman-break galaxies (LBGs).
To constrain their extinction and intrinsic SFR,
we combine the latest ultradeep VLA 1.5 GHz radio image and the \emph{HST}
ACS optical images in the GOODS-N. We select
a large sample of 1771 $z\sim4$ LBGs from the ACS catalogue using $\bband$-dropout color criteria. 
Our LBG samples have $\iband\sim25$--28 (AB), $\sim0$--3 magnitudes fainter than $M^\star_{\rm UV}$ at $z\sim4$.
In our stacked radio images, we find the LBGs to be point-like under our $2\arcsec$ angular resolution.
We measure their mean 1.5 GHz flux by stacking the measurements on the individual objects.  We achieve a statistical detection of 
$S_{1.5\,\rm GHz}=0.210\pm0.075$ $\mu$Jy at $\sim3~\sigma$, first time on such a faint LBG population at $z\sim4$. The measurement takes into account
the effects of source size and blending of multiple objects.  The detection is visually confirmed by stacking the radio 
images of the LBGs, and the uncertainty is quantified with Monte Carlo simulations on the radio image. 
The stacked radio flux corresponds to an obscured SFR of $16.0\pm5.7$ $\solarmassperyear$, and implies
a rest-frame UV extinction correction factor of 3.8. This extinction correction is in excellent agreement with that 
derived from the observed UV continuum spectral slope, using the local calibration of \citet{meurer99}.  
This result supports the use of the local calibration on high-redshift LBGs for deriving 
the extinction correction and SFR, and also disfavors a steep reddening curve such as that of the Small
Magellanic Cloud.
\end{abstract}

\section{Introduction}
Lyman-break galaxies (LBGs) are a key galaxy population for the studies of the evolution of galaxies and 
intergalactic medium (IGM) in the 
distant universe. The most prominent signature in their optical (rest-frame UV) spectral energy distribution (SED) 
is a strong Lyman continuum break \citep{cowie88,songaila90,madau95} caused by the absorption of neutral hydrogen within the 
galaxies and in the IGM.  By identifying the color break (aka.\ ``dropout'') between two adjacent filter bands, deep optical surveys 
were able to efficiently pickup $z>3$ galaxies (\citealp{steidel92,steidel93,steidel95,steidel99}; see \citealp{giavalisco02} and 
references therein).  Large spectroscopic surveys targeting on LBGs \citep[e.g.,][]{steidel03} or on flux-limited complete galaxy samples
\citep[e.g.,][]{barger08} have both confirmed the effectiveness of the LBG color selection technique.

In the past decade, near-infrared imaging with space-based \citep[e.g.,][]{bouwens04,yan04,bouwens05,oesch10,mclure10} and 
ground-based \citep[e.g.,][]{stanway03,ouchi09,capak11,hsieh12,bowler12,hathi12} instruments have 
extended the search of high-redshift LBGs to $z>6$.  To date, such rest-frame UV selected galaxies have served as a key 
tracer in the era of reionization for the studies of the ionization status of the IGM \citep{ono12,schenker12,treu13,caruana14,cassata14} 
and the source of ionizing photons \citep[e.g.,][]{bunker10,bouwens12a,finkelstein12b,robertson13}.

To characterize the LBG contributions to the cosmic star formation and the cosmic reionization, 
a key measurement is their star formation rates (SFRs).
The most commonly adopted LBG SFR estimate is the rest-frame UV (1500--2800 \AA) luminosity, a measure of the amount of massive young stars in a galaxy
\citep{cowie97,Madau98}. 
The major uncertainty in the UV SFR is extinction correction.  On lower redshift galaxies, robust determinations of extinction can be achieved by fitting
the rest-frame UV to optical SEDs of galaxies.  This becomes very challenging for galaxies at $z>4$,
because the rest-frame optical light redshifts to 2 $\mu$m and longer wavebands. To overcome this, many studies 
\citep[e.g.,][]{ouchi04,stanway05,hathi08,bouwens09} measure UV continuum spectral slopes ($\beta$, defined as $f_\lambda \propto \lambda^\beta$) 
of LBGs, and derive extinction by  assuming a relation between $\beta$ and extinction \citep[][hereafter M99]{steidel99,meurer99}.
In particular, the calibration of the $\beta$--extinction relation based on local starburst galaxies of M99
is very often adopted by studies of high-redshift LBGs.

The method based on $\beta$ is sensitive to the assumptions on the extinction curve and the 
intrinsic UV spectral slope of galaxies. These are related to the dust properties and the stellar population (metallicity and age),
and are suggested by models to evolve at high redshifts \citep{gonzales-perez13,wilkins13}.  
In the observational side, various recent studies of local star forming galaxies \citep[e.g.,][]{takeuchi10,takeuchi12,overzier11} 
and high-redshift UV selected galaxies \citep[e.g.,][]{buat12}
have found $\beta$--extinction relations deviating from that in M99.
There have also been inconsistencies on the measurements of $\beta$ for high-redshift LBGs
\citep{bouwens12b,castellano12,finkelstein12a,dunlop12,bouwens14}, and more generally, 
$L^\star$ galaxies at $z\gtrsim2$ \citep[e.g.,][]{reddy06,daddi07}. These all make the SFR measurements uncertain.

To verify the extinction correction and dust properties of high-redshift LBGs, independent measurements of SFRs are required.
Several attempts have been made in detecting large samples of LBGs in the radio \citep[e.g.,][hereafter Ho10]{reddy04,carilli08,ho10}, 
submillimeter \citep[e.g.,][]{peacock00,chapman00,webb03,davies13}, far-infrared \citep[e.g.,][]{rigopoulou10,reddy12a,lee12,oteo13,davies13}, 
and X-ray \citep[e.g.,][]{reddy04,lehmer05,cowie12,basu-zych13}, to estimate their SFRs without the effect of dust extinction.  
Not all of these observations confirm the M99 $\beta$--extinction relation on high-redshift LBGs.
Moreover, generally speaking, successful detections of LBGs at these wavebands are mostly limited to either lower redshifts ($z\lesssim3$) 
or the most luminous galaxies. 
There remains a lack of extinction-free SFR measurements for faint $z\gtrsim4$ LBGs whose luminosity is more typical and 
closer to the faint $z>6$ LBGs that may be responsible for the cosmic reionization.

In order to constrain the SFR of $z\sim4$ faint LBGs at wavebands free from dust extinction, we have performed radio stacking analyses
of large samples of LBGs selected with extremely deep optical imaging. Radio synchrotron emission from normal galaxies is generated
in supernova remnants and is thus an excellent tracer of star formation \citep{condon92}. It is not affected by dust extinction.  
In addition, radio interferometric imaging has the advantage of high angular resolution, so fluxes boosted by blending of objects and by 
the contribution from background confusing sources is much easier to estimate (e.g., Section~\ref{sec_flux_stack}).  In our previous 
study (Ho10), we stacked $\sim3500$ $B$-band
dropouts ($z\sim3.8$) at 1.4 GHz. The LBGs were selected with the \emph{Hubble Space Telescope} (\emph{HST}) Advanced Camera for Surveys
(ACS) imaging data of the Great Observatories Origins Deep Survey-North and South (GOODS-N and GOODS-S, \citealp{giavalisco04}).
With the deep radio images of \citet{miller08} and \citet{morrison10}, Ho10 did not detect the $z\sim3.8$ LBGs.
Here we improve the stacking analyses, with the latest ultradeep 1.5 GHz radio image of the GOODS-N made with the Karl G.\ Jansky Very Large Array (VLA) of the 
National Radio Astronomy Observatory.
This unprecedentedly deep radio image allows us to detect the faint LBGs and place constraints on their
SFR and dust extinction.

\addtocounter{footnote}{+3}

The paper is structured as following. We describe our radio imaging and LBG galaxy samples in Section~2.
We present our stacking analyses in image and flux domain, the uncertainty estimates, and the results in Section~3. 
We discuss the significance of the result, the derived SFR and extinction of the LBGs, and the implication on
the extinction curve in Section 4.  We summarize in Section 5.  Throughout the paper, we adopt a flat 
$\Lambda$CDM cosmology, with $\Omega_m = 0.315$, $\Omega_\Lambda = 0.685$, and $H_0 = 67.3$~km~s$^{-1}$~Mpc$^{-1}$
\citep{planck13}\footnote{Many previous studies assume $\Omega_m = 0.3$, $\Omega_\Lambda = 0.7$, and $H_0 = 70$~km~s$^{-1}$~Mpc$^{-1}$. 
Comparing to this, our assumed cosmology has a luminosity distance that is 2.4\% larger at the redshift of interest ($z\sim3.8$), leading to a 4.9\% 
larger luminosity.}.  All magnitudes are in the AB system.

\section{Data}
\subsection{1.5 GHz Radio Imaging}

The details of the radio observations and data reduction will be presented in F.\ Owen (2014, in prep.).  Here we provide a brief summary.
The GOODS-N field was observed with the VLA in the A configuration, for a total of 39 hours including calibration and move time, 
between August 9 and September 11, 2011. Eight different scheduling blocks were observed, each of 5 hours, except for
one which was 4 hours long.  Roughly 33 hours of this time were spent on-source. The observations covered
the bands from 1000--1512 and 1520--2032 MHz using 1 MHz channels. A phase, bandpass, and  instrumental polarization
calibrator, J1313+6735, was observed every twenty minutes. 3C286 was observed to calibrate the flux density scale. 

	For each scheduling block, the data were edited and calibrated in AIPS. The worst parts of the band, in 
particular between 1520 and 1648 MHz were flagged at the beginning of this process. The rest of the dataset was edited 
using the {\tt RFLAG} task. After total intensity calibration, the $uv$-data weights were calibrated using the AIPS task, 
{\tt REWAY}. 

	The total intensity data were imaged in CASA using the clean task. In particular, the wide-field, {\tt nterms=2}
parameters were used. For these parameters, the Multi-Scale-Multi-Frequency-Synthesis algorithm is used \citep{rau2011}. 
This imaging algorithm solves for the total intensity and spectral index image across the full 
bandwidth, i.e., in this case  1-2 GHz. The final image was corrected for the primary beam attenuation, which was calculated
and corrected using the CASA task {\tt widebandpbcor}. The final, full resolution image is $7000 \times 7000$ pixels of $0\farcs35$ 
with a clean beam of $1\farcs59\times1\farcs37$ at pa = $92\arcdeg$ and  has a central rms noise of 2.2 $\mu$Jy
(corresponding to an SFR of 167 $\solarmassperyear$ at $z=3.8$, see Section~\ref{sec_sfr}).  
For this work, the cleaned image is further convolved with a Gaussian to have a circular beam of $2\arcsec$.
This way, both cleaned bright point sources and uncleaned faint point sources would have comparable sizes in the image.
We have compared our astrometry with the previous deep VLA imaging of \citet{morrison10}. There is a $0\farcs21$ 
offset along the R.A.\ direction.  This offset is much smaller than our beam size and our pixel size, and thus has negligible 
impact on our analyses.

\subsection{Lyman-Break Galaxy Samples}
We directly adopt the $z\sim4$ LBG samples used in Ho10. The Ho10 $B$-dropout samples were selected from 
the GOODS v2.0 ACS source catalog based on the SExtractor \citep{bertin96} AUTO-aperture magnitudes with the following criteria:
\[\bband -  \vband  >1.1,\\\]
\[\bband-\vband > (\vband-\zband)+1.1,\\\]
\[\vband-\zband <1.6,\\\]
\[S/N(\vband)>5, \ \ \  {\rm and} \ \ S/N(\iband)>3.\]
The above criteria are adopted from \citet{beckwith06} and \citet{bouwens07}, and are well established in the literature.  
In addition, Ho10 also
excluded compact objects whose SExtractor stellarity indices are greater than 0.8 and $\iband < 26.5$
to prevent stellar contamination.  A total of 1778 $\bband$-dropouts are selected from the GOODS-N region.
They have redshifts between $z\sim3$ and $\sim4.5$, with a mean of $z\sim3.8$ \citep{bouwens07}.
The GOODS v2.0 catalog corrects the astrometry offset between the radio and optical frames,
so we directly adopt the ACS source positions.

\begin{figure}[h!]
\epsscale{1.0}
\plotone{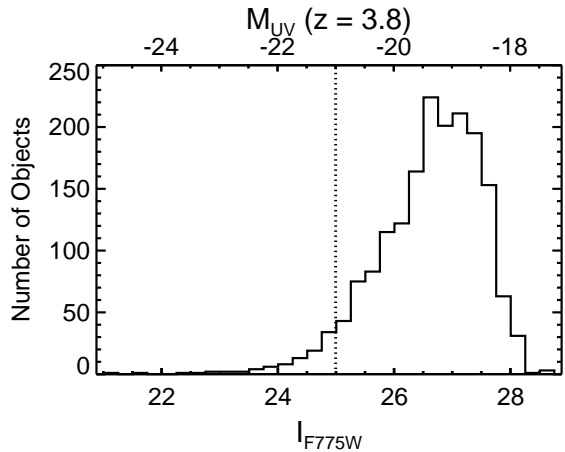}
\caption{Distribution of the $\iband$ magnitudes of our LBG samples and their absolute magnitudes assuming $z=3.8$.
The vertical dotted line indicates $M^\star_{\rm UV}$ of the $z\sim3.8$ LBG luminosity function.
\label{fig_mhisto}}
\end{figure}

We show the distribution of the $\iband$ magnitudes and the absolute UV magnitudes 
of our LBG samples in Fig.~\ref{fig_mhisto}.  The mean and median absolute magnitudes are $-19.40$ and $-19.27$, respectively.
The $B$-dropout luminosity function has a characteristic magnitude of $M^\star_{\rm UV}\sim-21.0$
\citep{bouwens07}. Therefore, our LBG samples are $\sim0$--3 magnitudes fainter than $M^\star_{\rm UV}$ and
represent the faint end of the $z\sim4$ LBG population.

\section{Stacking Analyses}
Measuring fluxes reliably in a radio image requires knowledge about the size of the targets. Ho10 assumed
that the size of the LBGs are similar to compact faint objects detected in their radio images.
In the present work, the high S/N offered by the new VLA image allows us to constrain the average size of the 
$z\sim4$ LBGs from the stacked image.  We first describe the image stacking and object size. Then we move to
more sophisticate, flux-based stacking analyses.

\subsection{Pre-Stacking in Image Domain}\label{sec_image_stack}

\begin{figure}[h!]
\epsscale{1.0}
\plotone{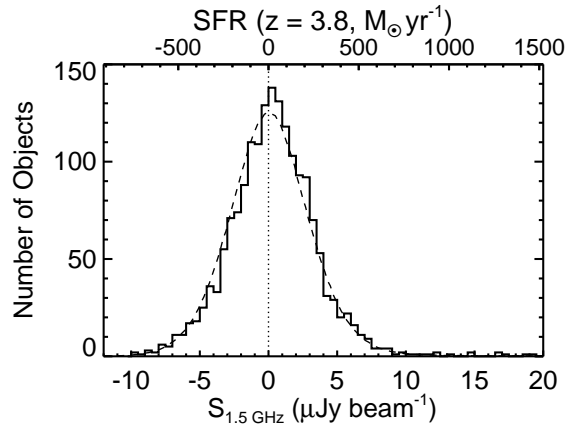}
\caption{Distribution of radio surface brightness measured at the optical positions of the LBGs (solid histogram) and
at random positions (dashed curve). 
There are  seven objects whose brightness are $>20$ $\mu$Jy beam$^{-1}$ and they are outside this plot.
The upper $x$-axis shows the corresponding SFR for objects at $z=3.8$.
\label{fig_fhisto}}
\end{figure}

We first measured the 1.5 GHz surface brightness of each LBG at its ACS position, in unit of Jy beam$^{-1}$ within
one $0\farcs35$ pixel.  The value would also be its total flux in Jy if it is a point source and if its radio position matches its ACS position.  
The distribution is 
presented Fig.~\ref{fig_fhisto} (solid histogram).  The majority of the objects (1771 out of 1778) have surface brightness between 
$-10$ $\mu$Jy beam$^{-1}$ and 20 $\mu$Jy beam$^{-1}$.  The $>20$ $\mu$Jy beam$^{-1}$ ones can be 
either intrinsically bright, or happen to fall on sight lines close to foreground radio-bright galaxies.
In the subsequent analyses, we primarily consider sources fainter than 20 $\mu$Jy beam$^{-1}$.  
This threshold corresponds to an SFR of $\sim1500$ $M_\sun$ yr$^{-1}$
at $z=3.8$, or an infrared luminosity of $L_{IR} \lesssim 10^{13}~L_\sun$ (Section~\ref{sec_sfr}).  
In other words, ULIRGs would be included in our analyses, but not the brighter submillimeter galaxies detected 
in ground-based surveys.  Increasing this upper threshold would make our 
results either more vulnerable to contamination from bright foreground galaxies and radio active galactic nuclei (AGNs), or being dominated 
by a small number of brighter LBGs.  Decreasing this threshold to close to 10 $\mu$Jy would significantly skew the statistics because
faint sources may be scattered by the Gaussian noise to this flux level.  We further discuss the effect of this threshold in Section~\ref{sec_threshold}.
In Fig.~\ref{fig_fhisto}, we also show the distribution of surface brightness at random positions within the GOODS-N region
(dashed curve).  Between $-10$ and $+10$ $\mu$Jy beam$^{-1}$, the histogram for the LBGs appears to shift positively
with respect to the histogram for random positions.  This indicates an excess of radio flux from the LBGs.

\begin{figure}[h!]
\epsscale{1.0}
\plotone{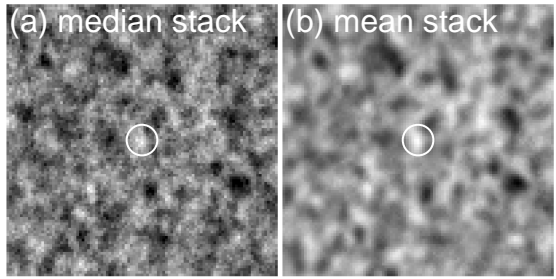}
\caption{Stacked radio images at the optical LBG positions.  The weighted median stack is shown in (a) and
the weighted mean stack is shown in (b).  The circles indicate the optical position and have radii of $2\arcsec$. \label{fig_stack_images}
The two images have identical brightness scales. North is up.}
\end{figure}

We next cut the radio image around the optical positions of the LBGs and stacked them.  In the stacking, we 
weighted the images with the inverse-square of the VLA primary beam response at the respective positions, which in principle reflects
the local noise level.  For our sources, the lowest primary beam response is 0.67, and $>60\%$ of the sources
are located in the region where the primary beam response is $>0.9$. 
Therefore, such a weighting scheme does not introduce a strong bias toward a small number of objects near 
the center of the primary beam.
We excluded images for LBGs that are brighter than 20 $\mu$Jy beam$^{-1}$.
For the images included in the stacking, we also excluded any pixels that are brighter than 20 $\mu$Jy beam$^{-1}$.
We performed both weighted median\footnote{For the weighted median, we assigned to each element
a $dx$ that is proportional to the weight.  We then integrated the $dx$ according to the sorted values of the elements until 
a percentile of 50\%.} and weighted mean in the stacking.  
The former provides the typical properties of the majority of the LBGs, while the latter provides the averaged contribution 
to the total star formation in the LBG population.  The results are presented in Fig.~\ref{fig_stack_images}.
In both the median and mean stacked images, a significant signal appears at the stacked optical position.
The point-source flux measured at the optical position is $0.235\pm0.083$ $\mu$Jy (median) and 
$0.237\pm0.072$ $\mu$Jy (mean).  In the mean-stacked case, if we stack at random positions, there is a 0.087\%
probability for the stacked flux to be higher than the above 0.237 $\mu$Jy. This probability corresponds 
to $\sim3.1~\sigma$ if the distribution is Gaussian, and this is consistent with the measured significance level.  
Furthermore, if we allow for a slightly extended object, the integrated flux measured
in a $3\arcsec$ box becomes 0.301 $\mu$Jy in the mean-stacked case.  These results show that there is 
a measurable ($\sim3$--$4~\sigma$) mean radio flux from the 1771 LBGs.

The key question here is the size and morphology of the LBGs, which would strongly affect how their total fluxes should be measured.
In the mean-stacked image (Fig.~\ref{fig_stack_images} (b)), there is a slight hint that the stacked object is north-south elongated. 
The same exists in the median-stacked image (Fig.~\ref{fig_stack_images} (a)) but is less apparent.
If the radio morphology of the LBGs is intrinsically elongated, however, the elongation should not be aligned and the stacked image 
should not have a preferred direction.  Therefore, this should be either an effect of noise, or some systematic effects.  
For the former, given that this elongation is less apparent in the median-stacked image, we think this 
may be an effect of a small number of noise spikes or nearby radio sources that fall close to the LBGs by chance projection.  We will further quantify 
the effect of chance projection in the next subsection.  A potential systematic effect that may produce the observed elongation is a 
minor north-south misalignment between the radio and optical astrometry whose offset varies across the GOODS-N field.
However, we carried out a simulation of perturbed astrometry and we found that an rms astrometry offset 
of $1\farcs4$ is required to produce the observed elongation.  Such an rms offset is unusually large and is therefore unlikely. 
Nevertheless, the above elongation is quite weak.  By fitting the image with Gaussian models, we obtained major-axis 
sizes of $3\farcs5\pm2\farcs0$ (median) and $4\farcs1\pm1\farcs7$  (mean), and minor-axis sizes of $2\farcs0$ in both cases.
All these values do not significantly deviate from the $2\arcsec$ beam size, meaning that the intrinsic radio size of the LBGs is compact
relative to our beam, and that the uncertainty in flux measurement introduced by the morphology/astrometry uncertainty is 
at most at a level of 1~$\sigma$.  In the subsequent analyses, we assume that the LBGs are all point sources under our resolution,
and we do not apply any correction to their optical positions.  Under such a conservative approach, our measurement may slightly
under-estimate the flux of LBGs for up to 1 $\sigma$.

\subsection{Flux Stacking}\label{sec_flux_stack}
We measured the radio fluxes of $z\sim4$ LBGs at their optical positions assuming point sources.
We excluded sources with radio fluxes $>20$ $\mu$Jy. For the remaining 1771 $<20$ $\mu$Jy sources,
we averaged their fluxes by weighting them with the inverse-square of the VLA primary beam 
response, and subtracted a background value to obtain the final stacked radio flux.
The uncertainty and the background of the above measurement were estimated with the following Monte Carlo 
simulations with the assumption that the spatial distribution of radio sources is random. 

We measured the 
radio fluxes at random positions in the radio image within the GOODS-N area, and excluded values higher than 20 $\mu$Jy. 
The number of random positions is the same as that of the $<20~\mu$Jy LBGs. Here we also weighted the fluxes with the inverse-square of the VLA primary beam response to obtain a mean value. The measurements were repeated $10^4$ times, and the distribution of the mean radio fluxes appears to be nearly Gaussian. 
The mean of the distribution was considered as a background and subtracted from the stacked LBG radio flux. This subtraction is to account for the 
contribution from confusing sources near our targets. We use the dispersion among the above $10^4$ Monte Carlo mean radio fluxes to represent 
the uncertainty of the stacked LBG radio flux. 

There is one important systematic effect in the above procedure. If multiple sources are blended in the image, then the stacked radio flux will be overestimated by our stacking method. Various deblending methods had been used in stacking low-resolution submillimeter data \citep[e.g.,][]{kurczynski10,greve10}.
However, in our case, the blending effect is weaker than that in the submillimeter, because our radio image has a much higher angular resolution. Here we adopt a simple approach. Our blending correction is only made on LBGs with one or more LBG neighbors within $3\arcsec$. Sources separated by more than $3\arcsec$ (three times the beam HWHM) have negligible effects on each other. We consider two types of blended objects in the radio image. One is sources with only one neighbor (262 objects), and the other is sources with two neighbors (30 objects). For the first type (pairs), we deblended them by assuming a Gaussian profile whose FWHM is identical to the beam FWHM, and corrected for the emission contributed by their neighbors. The correction is made on each pair of objects according to their angular separation. This method is the same as the one used by \citet{Webb04} . For the second type of groups, we corrected their stacked fluxes using simulations. We simulated such three-source groups by first placing one object at the image center and randomly placing the other two objects within $3\arcsec$ from the first one. The fluxes of the individual sources are random. All sources are point sources, convolved with the $2\arcsec$ telescope beam. We measured the fluxes at the source positions, and computed the ratio between the measured stacking flux and the total input flux. We repeated this by $10^6$ times and obtained an averaged ratio of 1.38. We divided the flux of sources in the second type of groups by this factor. We applied the above blending corrections both to the stacked radio flux and the Monte Carlo fluxes.

In the above deblending procedure, there is one subtle consideration. Although we applied the same correction to both the stacked radio flux and the Monte Carlo fluxes, the spatial distributions (i.e., clustering) of the Monte Carlo sources and the LBGs are not the same. To account for this, we performed another version of Monte Carlo simulations. Instead of picking up fluxes from random positions, we used the true positions of the $z\sim4$ LBGs and applied a random shift to all of them in each Monte Carlo run.  
We repeated this for a thousand runs with random offsets ranging from $10\arcsec$ to $40\arcsec$, and calculated their mean and dispersion. Because the relative positions of Monte Carlo sources are kept the same as those of the $z\sim4$ LBGs, we applied the same blending corrections to the blended Monte Carlo sources. Hereafter we refer the Monte Carlo simulations
based on random positions as ``MCA,'' and the Monte Carlo simulations based on randomly shifted real LBG positions as ``MCB.'' 

Finally, we compare fluxes from the isolated LBGs and the LBGs with $<3\arcsec$ neighbors.
The stacked, blending-corrected flux of the 292 LBGs with neighbors is $0.285\pm0.185~ \mu$Jy, which is slightly higher than
the flux of $0.193\pm0.082~ \mu$Jy for the 1479 isolated LGBs.  This marginal difference is consistent with the common expectation 
that galaxies undergoing merging or interaction have higher SFRs.  However, this $<1\sigma$ difference has to be confirmed with more 
sensitive radio imaging or larger LBG samples. 

\begin{deluxetable}{ccccc}
\tablecaption{Mean Radio Fluxes from Various Stacking Methods\label{tab1}}
\tablehead{ \colhead{Threshold} & \colhead{$N_{\rm LBGs}$} & \colhead{$S_{\rm MCA}$} & \colhead{$S_{\rm MCB}$} & \colhead{$S_{\rm IMG}$} \\
\cline{3-5}
\colhead{($\mu$Jy)} & & \multicolumn{3}{c}{($\mu$Jy)} 
}
\startdata
10  & 1757 & $0.166\pm0.072$ & $0.150\pm0.066$ &  $0.183\pm0.069$                                \\  
15  & 1766 & $0.186\pm0.074$ & $0.176\pm0.067$ & $0.208\pm0.071$ \\ 
20  & 1771 & $0.210\pm0.075$ & $0.202\pm0.070$ & $0.237\pm0.072$ \\  
25  & 1773 & $0.216\pm0.077$ & $0.206\pm0.070$ & $0.241\pm0.074$   
\enddata
\end{deluxetable}

\subsection{Results}
We summarize our stacking results in Table ~\ref{tab1}. Additional to results derived from excluding $>20$ $\mu$Jy sources, we also present
results with various upper flux thresholds.  Column (2) shows the numbers of the $z\sim4$ LBGs. Column (3) and column (4) are the stacked 
fluxes with uncertainties estimated by the two versions of Monte Carlo simulations.  Column (5) shows results from image stacking (no deblending) described 
in Section~\ref{sec_image_stack}.  The stacking fluxes derived with the the two versions of Monte Carlo simulations are very similar, given the 
error bars. In the subsequent analyses we adopt the results based on the $20$ $\mu$Jy threshold and MCA. 
We further discuss this choice of threshold in Section~\ref{sec_threshold}.
The stacked radio flux of $z\sim4$ LBGs is $2.8\sigma$. 
After all the attempts to systematically decrease the signal level 
(the point-source assumption, the subtraction of a mean background, and the 
downward corrections for the effects of blending), we consider this nearly $3\sigma$ result a robust detection.
This result is also consistent with that derived with image stacking (Section~\ref{sec_image_stack}).

In Ho10, a mean 1.4 GHz flux of $-0.05\pm0.18$ $\mu$Jy is measured from the same  LBG samples in the GOODS-N,
using the VLA image of \citet{morrison10} and an upper flux threshold of 100 $\mu$Jy. 
We repeated our stacking with the same 100 $\mu$Jy threshold and without weighting, and found a stacked 
flux of $0.298\pm0.088$ $\mu$Jy on 1777 sources. This is $<2~\sigma$ higher than the Ho10 value.
While the improvement here is substantial, the measured flux is still consistent with that in Ho10 given their uncertainty of 0.18 $\mu$Jy.


Prior to Ho10, \citet{carilli08} performed similar radio stacking analyses on $z\sim3$, 4, and 5 LBGs in the COSMOS field.
From their 1447 $z\sim4$ $B$-dropout samples, they obtained a 2 $\sigma$ detection of a median flux of $0.83\pm0.42$ $\mu$Jy.
Their stacked flux is much higher than ours, likely because their sources are much more luminous.
The LBG selection in Carilli et al.\ was based on the first data release of COSMOS \citep{capak07}, which has a $V$-band limiting 
magnitude of 25.0, much shallower than our limiting magnitude of $\vband>28$ (also see Fig.~\ref{fig_mhisto}).  
For a better comparison, we repeated our stacking analyses on the 41 $\vband<25.0$ sources in our ACS sample
without applying any radio flux threshold. We obtained a median flux of $0.63\pm0.57$ $\mu$Jy, consistent with the
result in \citet{carilli08}.  We conclude that we have obtained the first radio detection of faint $z\sim4$ LBGs,
but the GOODS-N area coverage is less optimal for studying the more luminous $z\sim4$ LBGs in the radio.

\section{Discussion and Implication}
\subsection{Radio-Bright Objects and the Stacking Threshold}\label{sec_threshold}
\begin{figure}[h!]
\epsscale{0.8}
\plotone{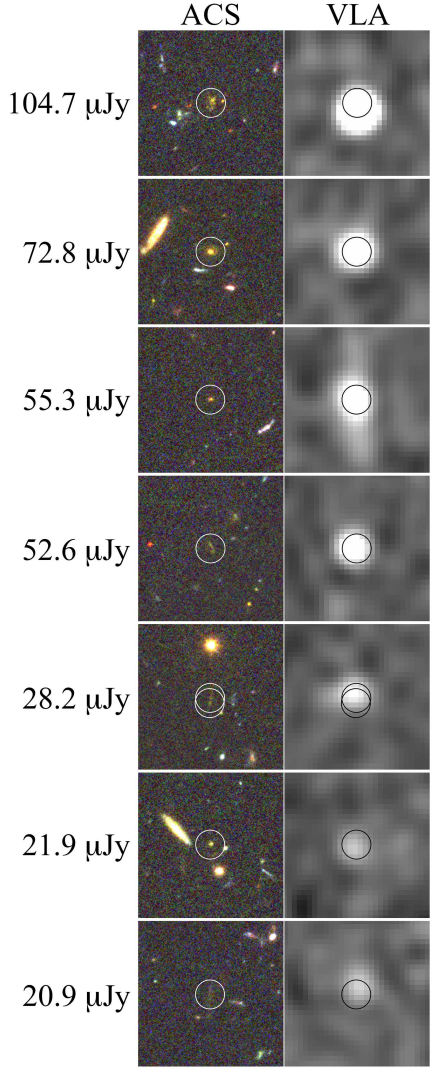}
\caption{The seven brightest LBGs. The grayscale pictures in the right column are
the VLA 1.5 GHz images. The three-color pictures in the left column are composed with the \emph{HST} ACS
$\bband$ (blue), $\vband$ (green), and $\iband+\zband$ (red) images. Under such a color scheme, $\bband$-dropout
objects have yellow colors.  Circles indicate the optical positions of the LBGs, and have radii of $1\arcsec$. 
The fifth galaxy is listed as two separated sources in the GOODS-N catalog, and both sources satisfy our LBG selection criteria.
These seven LBGs are excluded in our stacking analyses to avoid radio AGNs and effect of bright sources.
The remaining 1771 LBGs appear to be comparable with noise in the radio image.
\label{thumbnails}}
\end{figure}

Although the $z\sim4$ ACS-selected LBGs are radio-faint on average, there are a few radio-bright objects.
In Fig.~\ref{thumbnails} we show the seven $>20$ $\mu$Jy objects.  
The first four $>50$ $\mu$Jy ones would
have SFRs of $\gtrsim4000~\solarmassperyear$, if their radio emission were powered by star formation. 
However,  such high SFRs are close to or exceed the maximum SFR suggested in the radio studies of 
submillimeter sources in the GOODS-N \citep{barger14}, and these sources  
do not appear in the latest deep AzTEC \citep{perera08} and SCUBA-2 \citep{barger14} images
in the millimeter and submillimeter (mm/submm).  Therefore, their strong radio emission should contain substantial
contributions from radio AGNs.
One particularly interesting object is the 104.7 $\mu$Jy source, the brightest one in our LBG samples.
It has an extremely blue rest-frame UV continuum, with a spectral slope of $\beta=-3.86$ 
(see next subsection). This slope cannot be produced by any young stellar populations, and 
thus implies an unobscured AGN. On the other hand, this source is also qualified as an extremely red object
at the $K_S$ and IRAC bands \citep[KIERO,][]{wang12}. So this galaxy is a combination of an unobscured radio AGN and
a massive (or dusty) host galaxy. We do not include this object in our subsequent analyses, but we include the
remaining six $>20$ $\mu$Jy sources in the discussion in Section~\ref{sec_sfr}.

In the 10--30 $\mu$Jy range, objects with cooler dust remain above the sensitivity limits
of mm/submm surveys.  Three of the 16 sources between 10 and 30 $\mu$Jy are identified as 
mm/submm sources (AzGN21 in Perera et al.\ 2008, and CDFN30 and CDFN31 in Barger et al.\ 2014).
The mm/submm waveband primarily probes emission from cool dust in star-forming regions (cf. warm/hot 
dust around AGNs).  The radio-to-mm/submm flux ratios of these three sources are also broadly consistent
with that of dusty star-forming galaxies \citep[e.g.,][]{carilli99,barger00}.  Therefore, the radio and dust emission from
these three galaxies is most likely to be predominated powered by star formation.
The same may apply to the other 10--30 $\mu$Jy sources.  Their expected mm/submm fluxes 
(based on their radio fluxes and assuming dusty star formation) are close to current single-dish detection limits, so 
non-detections in the mm/submm surveys do not rule out cool dust emission.
The above observations suggest that once we exclude sources brighter than roughly 30 $\mu$Jy, 
we are in the star forming galaxy regime, although we cannot entirely rule out contribution from radio AGNs.
All the $>30$ $\mu$Jy sources are excluded in our stacking analyses.

Below the above mentioned $\sim30$ $\mu$Jy level, it is possible to set an even lower flux threshold to
prevent our stacking results from being dominated by a few bright objects.  On the other hand,  as mentioned in 
Section~\ref{sec_image_stack}, having a threshold close to 10 $\mu$Jy would also significantly skew
the results and under-estimate the mean flux.  Between the two extreme cases (30 and 10 $\mu$Jy),
we see in Table~\ref{tab1} that decreasing this threshold progressively decreases the stacked mean 
flux, flux uncertainty, and S/N.  The smooth behavior here indicates that the signal from $<10$ $\mu$Jy
sources is likely to be real, and that the underlying population has a continuous contribution to the measured radio
flux in the 10--30 $\mu$Jy interval.  Therefore, the choice of any threshold listed in Table~\ref{tab1} is 
just a matter of how many brighter objects that we feel comfortable to throw away.  
Adopting a different threshold would change the signal at $\lesssim1\sigma$ level, and only
slightly affects how we interpret the results.  In the present work, we adopt the results with a 20 $\mu$Jy threshold.  
Comparing to the dispersion of the fluxes at random positions (dashed curve in Fig.~\ref{fig_fhisto}), 
this is 7.0 $\sigma$.  So we are only discarding objects that are most securely detected.

\subsection{Star Formation Rate and Extinction}\label{sec_sfr}
We estimate the SFR of the LBGs using the stacked 1.5 GHz flux.  This is done by calibrating the SFR through
the widely adopted conversion between SFR and infrared luminosity of galaxies, and the local radio--FIR correlation, 
which is traditionally defined on the rest-frame 1.4 GHz radio power.
First, the rest-frame 1.4 GHz power can be expressed as
\begin{equation}
L_{\rm 1.4\,GHz}=4\pi D_l^{2}S_{\rm 1.4\, GHz}(1+z)^{-(1+\alpha)},
\end{equation}
where $D_l$ is the luminosity distance, and we assume a universal synchrotron  spectral index of $\alpha=-0.8$
($\alpha$ defined as $f_\nu \propto \nu^\alpha$).
For local normal star-forming galaxies, the radio power is proportional to the FIR luminosity \citep{condon92}:
\begin{equation}
q = \log\frac{L_{\rm FIR(40-120\,\mu m)}}{3.75\times10^{12}\,\rm W} -  \log\frac{L_{\rm 1.4\,GHz}}{\rm W\,Hz^{-1}},
\end{equation}
where $q$ is 2.3 on average. 
We adopt the conversion between total infrared luminosity $L_{\rm IR(8-1000\,\mu m)}$ and SFR in \citet{Kennicutt98}:
\begin{equation}
\frac{\rm SFR}{\solarmassperyear} = 4.5\times10^{-44} 
~\frac{L_{\rm IR(8-1000\,\mu m)}}{\rm erg\ s^{-1}},
\end{equation}
where a Salpeter initial mass function is assumed.
The ratio of $L_{\rm IR(8-1000\,\mu m)}/L_{\rm FIR(40-120\,\mu m)}$ is roughly 2.0, and this is valid for a broad range of dust properties. 
The combination of the above equations provides
\begin{equation}
\frac{\rm SFR}{\solarmassperyear} = 7.38\times10^{-22} 
~\frac{L_{\rm1.4\,GHz}}{\rm W\,Hz^{-1}}.
\end{equation}
For $z=3.8$ and our observed waveband of 1.5 GHz (converting to 1.4 GHz using $\alpha=-0.8$), the final SFR formula is then
\begin{equation}
\frac{\rm SFR}{\solarmassperyear} = 76.1
~\frac{S_{\rm1.5\,GHz}}{\rm \mu Jy}.
\end{equation}
The SFR of our $z\sim4$ LBGs estimated with our stacked radio flux and the MCA uncertainty is thus $16.0\pm5.7~\solarmassperyear$. 

We can also estimate the SFR using the fluxes in the $\iband$ and $\zband$ bands, which correspond to the rest-frame UV. 
We adopt the calibration in \citet{Kennicutt98}:
\begin{equation}
\frac{\rm SFR}{\solarmassperyear} = 1.4 \times 10^{-28} \frac{L_{\rm UV}} {\rm (erg\ s^{-1} Hz^{-1})},
\end{equation}
where $L_{\rm UV}$ is the rest-frame UV luminosity density between 1500 \AA\, and 2800 \AA, and a Salpeter initial mass function is assumed.
We calculated the mean $\iband$ and $\zband$ fluxes of the 1771 $S_{\rm 1.5\, GHz}<20$ $\mu$Jy LBGs by applying the same weighting
factors in the radio stacking analyses based on the VLA primary beam response.
The results are 0.133 $\mu$Jy for $\iband$ (26.090 AB) and 0.142 $\mu$Jy for $\zband$ (26.019 AB), respectively,
corresponding to SFRs of 5.48 $\solarmassperyear$ and 5.85 $\solarmassperyear$, uncorrected for extinction.  

The above radio and UV SFRs represent obscured and unobscured star formation, respectively. 
The total star formation can be expressed as the sum of the two (e.g., \citealp{reddy12a}, see \citealp{Kennicutt12} for more complete
discussion on composite SFR). Therefore, the total (radio+UV) SFR of our LBGs is 21.5--21.9 ($\pm$5.7) $\solarmassperyear$.
The ratio between the above total SFR and the uncorrected UV SFR thus implies an extinction correction of  roughly 3.8 ($\pm1.0$).

In the studies of high-redshift LBGs, the spectral slope in the UV continuum ($\beta$) is often used for estimating extinction.
This is because full optical-to-near-infrared SED fitting for extinction and mid/far-infrared detection of the obscured star formation 
are both challenging for faint LBGs.  Here we assume the correlation
between $\beta$ and extinction on local starbursting galaxies (M99):
\begin{equation}\label{eq7}
A_{1600}\, ({\rm mag}) = 4.43 +1.99\beta,
\end{equation}
where $A_{1600}$ is the extinction at rest-frame 1600 \AA, and $\beta$ is the spectral slope between 1300 \AA\, and 2600 \AA. With the mean 
$\iband$ and $\zband$ magnitudes (converted from the mean fluxes), we compute the UV spectra slope $\beta$ with the following relation \citep{bouwens09}:
\begin{equation}
\beta^\prime = 5.30(\iband-\zband)-2.04,
\end{equation}
\begin{equation}
\beta=-2.31+1.11(\beta^\prime+2.3),
\end{equation}
where $\beta^\prime$ is the continuum slope across 1600 to 2300 \AA, and $\iband$ and $\zband$ are AB magnitudes. 
With the above equations we obtain a mean $\beta$ of $-1.604$ and a mean extinction of 
$A_{1600} \sim 1.24$ mag, which corresponds to an attenuation of $3.13\times$. 
This value agrees excellently with the extinction correction estimated with the radio and UV SFRs, which is $\sim3.8$ with a 26\% uncertainty.
This implies that there do not exist substantial star forming components in these faint LBGs that are completed obscured 
\citep[cf.][]{capak08,wang09}.  This also
supports the method of using $\beta$ and Eq.~\ref{eq7} to estimate the extinction in high-redshift LBGs for correcting their UV SFRs.

\begin{figure}[h!]
\epsscale{1.0}
\plotone{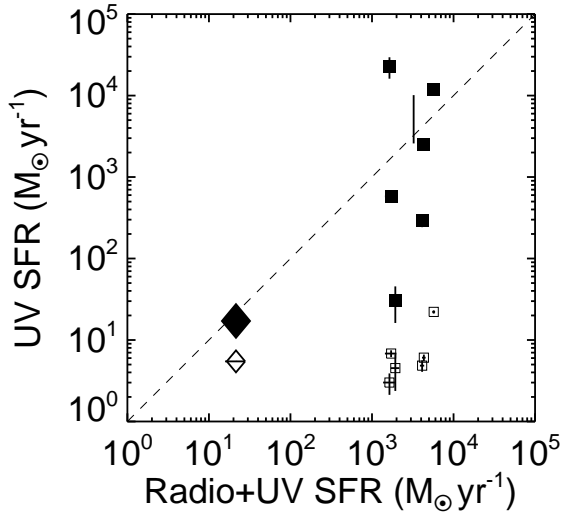}
\caption{UV SFR vs.\ total SFR (radio plus uncorrected UV). The squares are the six 20--100 $\mu$Jy LBGs, and the diamonds represent the
stacked LBG sample.  The solid symbols are extinction-corrected UV SFRs based on the UV spectral slope ($\beta$),
and the open symbols are uncorrected values. The solid vertical line shows the mean extinction-corrected UV SFR
for the 20--100 $\mu$Jy LBGs, and its length is the dispersion of the data divided by the square-root of the
source number.
\label{radio_uv_sfr}}
\end{figure}

In Fig.~\ref{radio_uv_sfr}, we present a comparison between the total SFR and UV SFR, for our stacked
LBG sample (diamonds), and the sub-sample of 20--100 $\mu$Jy sources (squares).  For the 20--100 $\mu$Jy
sources, the extinction-uncorrected SFRs (open squares) are all $\lesssim10~\solarmassperyear$, nearly two orders of magnitude
less than their radio SFRs. The above $\beta$-based extinction correction (solid squares) increases their SFR substantially, 
but the corrected UV SFRs have a spread of nearly three orders of magnitude, much larger than the spread in their total SFRs.
This suggests that the uncertainty of the $\beta$-based
extinction correction for the most intense starbursts can go either way. It can lead to an underestimate, most likely because of
the existence of completely obscured star formation. It can also lead to an overestimate, perhaps because of the
existence of established stellar populations that contaminate the UV SED. The possible existence of AGNs further complicates
the interpretations of both the radio and the UV emission. This result shows that  
SFRs derived with optical photometry and the $\beta$-based extinction correction should be treated with caution,
at least for the most intense star-forming galaxies.

An interesting question to ask here is, whether the $\beta$-based extinction correction also produces such 
a huge scatter on fainter LBGs? Naively speaking, LBGs with SFRs of $\lesssim100~\solarmassperyear$ should be less
massive and less dusty, and thus have less extinction.  This is supported by many previous studies 
\citep[e.g.,][]{wang96,burgarella09,reddy10} and the moderate extinction correction ($\sim3$) inferred for our stacked LBG sample.  
On the other hand, it is interesting
to note that even for the $>20$ $\mu$Jy objects, the mean of their extinction-corrected UV SFR (the vertical line in
Fig.~\ref{radio_uv_sfr}) also agrees with their mean total SFR.  Therefore, although the mean total SFR and the 
mean extinction-corrected UV SFR of the fainter LBGs agree well, this does not rule out a large internal scatter
in the $\beta$-based extinction correction.  Our data only show that the extinction correction leads to a correct SFR,
\emph{on average}, but not on individual objects.

\subsection{Shape of the Extinction Curve}

The $\beta$--extinction relation (e.g., Eq.~\ref{eq7}) contains the effects of both the
reddening curve and the intrinsic (unreddened) spectral slope of the stellar population.  
The locally calibrated Eq.~\ref{eq7} from M99 implies an intrinsic spectral slope of $\beta_{\rm int}=-2.23$ and a 
reddening curve that is similar to that in \citet{calzetti00}, but should not be considered unique.  For example, 
\citet{castellano14} demonstrated that bright LBGs at $z\sim3$ have lower metallicity and require a
different $\beta$--extinction relation.
\citet{wilkins13} employed semi-analytical galaxy formation models to calculate the
intrinsic UV spectral slopes of $z>5$ galaxies and found bluer UV continua of $\beta_{\rm int} \sim -2.4$ ($z\sim5$) up to 
$\gtrsim-2.7$ ($z\sim10$).  To quantitatively test this, we consider the expression in \citet{wilkins13}:
\begin{equation}
A_\lambda = D_\lambda \times [\beta_{\rm obs} - \beta_{\rm int}],
\end{equation}
where $\beta_{\rm obs}$ and $\beta_{\rm int}$ are the observed (reddened) and intrinsic UV spectral slopes, respectively. 
The factor $D_\lambda$ depends on the reddening curve. Wilkins et al. found $D_\lambda=1.84$, 0.96, 1.47, and 1.90,
for the reddening curves of \citet{calzetti00}, the Small Magellanic Cloud \citep[SMC,][]{pei92}, the Large Magellanic Cloud \citep[LMC,][]{pei92}, 
and core collapse supernovae \citep[CCSN,][]{bianchi07}, respectively, for $\lambda=1300$--2100 \AA.  Our measurements
provide $\beta_{\rm obs}=-1.604$ and $A_{1600}=1.12$.  For the above four reddening curves, the inferred intrinsic 
UV spectral slopes for the stellar population are then $\beta_{\rm int} = -2.21$, $-2.77$, $-2.37$, and $-2.19$, respectively.
The uncertainties are between 0.15 (Calzetti and supernovae) and 0.3 (SMC).  
Among the four reddening curves, the steepest SMC one is strongly disfavored by our data, as it requires an unusually blue UV continuum, 
much bluer than any $z\sim5$ samples presented in \citet{wilkins13}.  
We note that if we adopt the most aggressive stacked radio flux of $\sim0.3$ $\mu$Jy in Section~\ref{sec_image_stack} (i.e., no blending correction
and assuming extended sources), the inferred $\beta_{\rm int}$ would be an extremely blue $-3.18$ for the SMC case, which is even less likely.

\begin{figure}[h!]
\epsscale{1.0}
\plotone{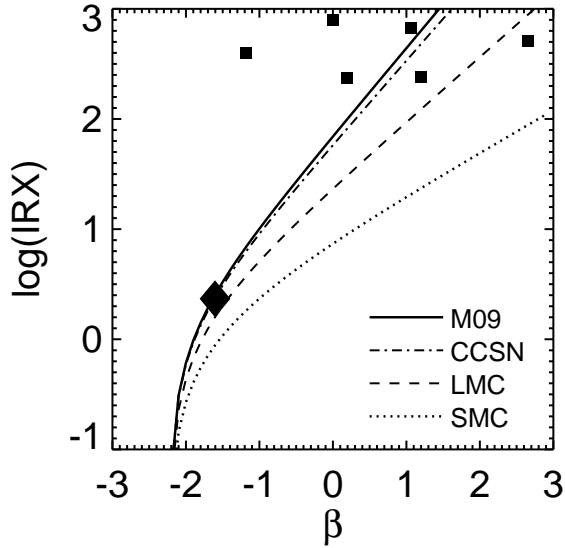}
\caption{$IRX$ vs.\ $\beta$ for the 20--100 $\mu$Jy LBGs (squares) and the stacked 1771 LBGs (diamond).
The solid curve is Eq.~\ref{eq7}, the local calibration of M99.  The dotted, dashed, and dash-dotted
curves show the reddening curves of SMC, LMC, and CCSN, respectively, coupled with $\beta_{\rm int}=-2.23$.
\label{beta_irx}}
\end{figure}

The above results based on Fig.~\ref{radio_uv_sfr} can be also presented as Fig.~\ref{beta_irx}, which is the diagram
that M99 used to present Eq.~\ref{eq7} (their Fig.~1).  The quantity $IRX$ is defined as $L_{\rm FIR}/(\nu L_{\rm UV})$, where
$L_{FIR}$ is the same 40--120 $\mu$m luminosity discussed in Section~\ref{sec_sfr} (also see \citealp{helou85}), and 
$L_{\rm UV}$ is the rest-frame 1600 \AA\, luminosity density. After taking into account all the conversion factors in 
Section~\ref{sec_sfr}, we find $IRX$ is approximately $0.83 \times SFR_{\rm radio}/SFR_{\rm UV}$. 
Fig.~\ref{beta_irx} shows $\beta$ and $IRX$ based on our UV and radio SFRs, for our stacked 1771 LBGs (diamond) and
the six 20--100 $\mu$Jy sources (squares). The local calibration of M99 is shown as the solid curve. The other three curves
shows the above mentioned SMC, LMC, and CCSN reddening laws, coupled with an assumed $\beta_{\rm int}=-2.23$.
Changing the assumed $\beta_{\rm int}$ will change where the curves intersect with the $x$-axis at $\log(IRX)=-\infty$, and only affects the
bottom part of the curves. Here we see again that the SMC extinction curve is not favored by our data, both for the
intense star forming galaxies and the stacked LBG.  

It is important to point out that latest studies of the UV spectral slopes of high-redshift LBGs have made use of 
more rigorous methods to measure $\beta$ \citep{bouwens12b,castellano12,finkelstein12a,dunlop12,bouwens14}. 
These studies tend to find smaller (bluer) $\beta$ values comparing to the simple two-band method \citep{bouwens09} 
adopted here.  Therefore, it is very likely that our $\beta$ values are slightly overestimated. However, from Fig.~\ref{beta_irx},
it is clear that a smaller $\beta$ makes the SMC extinction curve even more unfavorable.  So our conclusion is not affected 
by how $\beta$ is measured.  As mentioned above, this conclusion is also not affected by the conservative approach in our
radio stacking analyses, where we adopted a lower value for the stacked radio flux.

Extinction law in high-redshift star-forming galaxies is an unsettled issue.  In the literature, some studies favor an SMC-like extinction curve for 
$z>2$ galaxies (including LBGs), based on either rest-frame UV/optical SED fitting \citep[e.g.,][]{vijh03,verma07,oesch13} 
or the comparison between the infrared and UV luminosities \citep[e.g.,][]{reddy10,reddy12a,lee12}.  
On the other hand, latest \emph{Herschel} results of \citet{sklias14} on a sample of lensed star-forming galaxies
favor an extinction curve that is somewhat similar to the \citet{calzetti00} one, rather than an SMC-like extinction curve.
\citet{oesch13} also pointed out the inconsistent luminosity dependencies in the literature about whether an SMC or Calzetti dust is more 
suitable to the bright end or the faint end (see discussion therein).
Our result shows that in the $\sim10~\solarmassperyear$ low-luminosity end, an SMC-like extinction cannot explain the observed UV and radio SFRs. 
This probably adds more controversy to the current situation.  We thus conclude that the shape of extinction curve in high-redshift galaxies
and its dependency on galaxy populations remain open issues.

\section{Summary and Final Remark}
We select 1771 faint $z\sim4$ LBGs from the GOODS-N \emph{HST} ACS catalog and studied their averaged radio properties
with an ultradeep VLA 1.5 GHz image.  In our stacked images, we found the radio emission from the LBGs is on-average compact
under our $2\arcsec$ resolution.
We achieved a statistical detection of a mean radio flux of $0.210\pm0.075~\mu$Jy
with stacking analyses.  This radio flux corresponds to a mean obscured SFR of $16.0\pm5.7~\solarmassperyear$, which is $2.8\times$
higher than the unobscured SFR derived from the UV continuum of the LBGs. The ratio between the total (radio+UV) and UV SFRs (3.8) 
is in excellent agreement with the extinction inferred from the UV spectral slope.  
This also suggests an extinction curve that is similar  to that of local starburst galaxies in 
\citet{meurer99} and \citet{calzetti00}, instead of an SMC-like extinction curve.

In this work, we present the first radio detection of faint $z\sim4$ LBGs using a radio image taken during the very early 
phase of the Karl G.\ Jansky Very Large Array. However, it will be hard to push this to higher redshift in the near future even with
a deeper VLA image. From $z=4$ to $z=5$, the radio flux decreases by a factor of 1.4 to 1.6 (for $\alpha=0$ to $-0.8$, depending
on whether the rest-frame 10 GHz emission is synchrotron or free-free). This requires $>2\times$ of
observing time ($\sim100$ hr) on the VLA to achieve the same sensitivity per source.
Making this worse is the lack of an equally large ($>10^3$) sample of $z\sim5$ LBGs for  the same
luminosity (down to 3 magnitudes fainter than $M^\star_{\rm UV}$) for deep stacking analyses.  
One possible route is to obtain deep VLA images of lensing cluster fields such as the \emph{Hubble} Frontier Fields.
The strong lensing and the deep optical images may provide the needed radio sensitivity and the LBG samples.
On the other hand, it is also possible to place constraints on the obscured star formation using mm/submm 
observations instead.  In these wavebands, the dust spectral slope
produces a strong negative $K$-correction, making detections of high-redshift galaxies relatively easy.
The mean SFR of our LBGs corresponds to $L_{\rm IR}\sim10^{11}~L_\sun$.
This translates to fluxes of $10^{1.5-2}~\mu$Jy at 850 $\mu$m over a broad range of redshifts ($z\sim1$ to $\gtrsim6$).
Detecting such sources can be achieved with ALMA with stacking analyses on a relatively small LBG sample,
or even on individual LBGs.  We expect new constraints on the intrinsic SFRs of $z>4$ LBGs will soon be provided
by deep ALMA imaging in the GOODS-S, the \emph{Hubble} Ultra Deep Field, or similar \emph{HST} deep fields.

\acknowledgments
We thank H.\ Hirashita for useful discussion, and the referee for the thorough review. CHT and WHW are partially supported by the 
National Science Council of Taiwan grant 102-2119-M-001-007-MY3.  The National Radio Astronomy Observatory 
is a facility of the National Science Foundation operated under cooperative agreement by Associated Universities, Inc.


\begin{thebibliography}{} 
\bibitem[Barger, Cowie, \& Richards(2000)]{barger00}
	Barger, A.\ J., Cowie, L.\ L., \& Richards, E.\ A.\ 2000, \aj, 119, 2092
\bibitem[Barger, Cowie, \& Wang(2008)]{barger08}
	Barger, A.\ J., Cowie, L.\ L., \& Wang, W.-H.\ 2008, \apj, 689, 687
\bibitem[Barger et al.(2014)]{barger14}
   	Barger, A.\ J., Cowie, L.\ L., Chen, C.-C., et al.\  2014, \apj, 784, 9
\bibitem[Basu-Zych et al.(2013)]{basu-zych13}
	Basu-Zych, A.\ R., Lehmer, B.\ D., Hornschemeier, A.\ E., et al.\ 2013, \apj, 762, 45
\bibitem[Bertin \& Arnouts(1996)]{bertin96}
   	Bertin E., \& Arnouts S.\ 1996, \aaps, 117, 393
\bibitem[Beckwith et al.(2006)]{beckwith06}
   	Beckwith, S. V. W., et al 2006, AJ, 132, 1729
\bibitem[Bianchi \& Schneider(2007)]{bianchi07}
   	Bianchi, S., \& Schneider, R.\ 2007, \mnras, 378, 973
\bibitem[Bouwens et al.(2004)]{bouwens04}
   	Bouwens, R. J., Thompson, R.\ I., Illingworth, G., D., et al.\ 2004, \apj, 616, L79
\bibitem[Bouwens et al.(2005)]{bouwens05}
   	Bouwens, R. J., Illingworth, G., D., Thompson, R.\ I., Franx, M. et al.\ 2005, \apj, 624, 5
\bibitem[Bouwens et al.(2007)]{bouwens07}
   	Bouwens, R. J., Illingworth, G., D., Franx, M., \& Ford, H. 2007, \apj, 670, 928
\bibitem[Bouwens et al.(2009)]{bouwens09}
	Bouwens, R.\ J., et al. 2009, \apj, 705, 936
\bibitem[Bouwens et al.(2012a)]{bouwens12a}
	Bouwens, R.\ J., Illingworth, G.\ D., Oesch, P.\ A., et al.\ 2012a, \apj, 752, L5
\bibitem[Bouwens et al.(2012b)]{bouwens12b}
	Bouwens, R.\ J., Illingworth, G.\ D., Oesch, P.\ A., et al.\ 2012b, \apj, 754, 83
\bibitem[Bouwens et al.(2014)]{bouwens14}
	Bouwens, R.\ J., Illingworth, G.\ D., Oesch, P.\ A., et al.\ 2014, \apj, submitted (arXiv:1306.2950)
\bibitem[Bowler et al.(2012)]{bowler12}
	Bowler, R.\ A.\ A., Dunlop, J.\ S., McLure, R.\ J., et al.\ 2012, \mnras, 426, 2772
\bibitem[Buat et al.(2012)]{buat12}
	Buat, V., Noll, S., Burgarella, D., et al.\ 2012, \aap, 545, A141
\bibitem[Bunker et al.(2010)]{bunker10}
	Bunker, A.\ J., Wilkins, S., Ellis, R.\ S., et al.\ 2010, \mnras, 409, 855
\bibitem[Burgarella et al.(2009)]{burgarella09}
	Burgarella, D., Buat, V., Takeuchi, T.\ T., Wada, T., \& Pearson, C.\ 2009, \pasj, 61, 177
\bibitem[Calzetti et al.(2000)]{calzetti00}
	Calzetti, D., Armus, L., Bohlin, R.\ C., et al.\ 2000, \apj, 533, 682
\bibitem[Capak et al.(2007)]{capak07}
	Capak, P., Aussel, H., Ajiki, M., et al.\ 2007, \apjs, 172, 99
\bibitem[Capak et al.(2008)]{capak08}
	Capak, P., Carilli, C.\ L., Lee, N., et al.\ 2008, \apjl, 681, L53
\bibitem[Capak et al.(2011)]{capak11}
	Capak, P., Mobasher, B., Scoville, N.\ Z., et al.\ 2011, \apj, 730, 68
\bibitem[Carilli \& Yun(1999)]{carilli99}
	Carilli, C.\ L., \& Yun, M.\ S.\ 1999, \apjl, 513, L13
\bibitem[Carilli et al.(2008)]{carilli08}
 	Carilli, C.\ L., Lee, N., Capak P., et al.\ 2008, \apj, 689, 883
\bibitem[Caruana et al.(2014)]{caruana14}
	Caruana, J., Bunker, A.\ J., Wilkins, S.\ M., et al.\ 2014, \mnras, submitted (arXiv:1311.0057)
\bibitem[Cassata et al.(2014)]{cassata14}
	Cassata, P., Tasca, L.\ A.\ M., Le Fevre, O., et al.\ 2014, \aap, submitted (arXiv:1403.3693)
\bibitem[Castellano et al.(2012)]{castellano12}
	Castellano, M., Fontana, A., Grazian, L., et al.\ 2012, \aap, 540, A39
\bibitem[Castellano et al.(2014)]{castellano14}
	Castellano, M., Sommariva, V., Fontana, A., et al.\ 2014, \aap, in press (arXiv:1403.0743)
\bibitem[Chapman et al.(2000)]{chapman00}
	Chapman, S.\ C., et al.\ 2000, \mnras, 319, 318
\bibitem[Condon(1992)]{condon92}
	Condon, J.\ J.\ 1992, \araa , 30, 575
\bibitem[Cowie et al.(1988)]{cowie88} 
  	Cowie, L.\ L., Lilly, S.\ J., Gardner, J., \& Mclean, I.\ S.\ 1988, \apj, 332, L29
\bibitem[Cowie et al.(1997)]{cowie97}
	Cowie, L.\ L., Hu, E.\ M., Songaila, A., \& Egami, E.\ 1997, \apj, 481, L9
\bibitem[Cowie, Barger, \& Harsinger(2012)]{cowie12}
	Cowie, L.\ L., Barger, A.\ J., \& Harsinger, G.\ 2012, \apj, 748, 50
\bibitem[Daddi, et al.(2007)]{daddi07}
	Daddi, E., Dickson, M., \& Morrison, G., et al.\ 2007, \apj, 670, 156
\bibitem[Davies et al.(2013)]{davies13}
	Davies, L.\ J.\ M., Bremer, M.\ N., Stanway, E.\ R., \& Lehnert, M.\ D.\ 2013, \mnras, 433, 2588
\bibitem[Dunlop et al.(2012)]{dunlop12}
	Dunlop, J.\ S., McLure, R.\ J., Robertson, B.\ E., et al.\ 2012, \mnras, 420, 901
\bibitem[Finkelstein et al.(2012a)]{finkelstein12a}
	Finkelstein, S.\ L., Papovich, C., Brett, S., et al.\ 2012a, \apj, 756, 164
\bibitem[Finkelstein et al.(2012b)]{finkelstein12b}
	Finkelstein, S.\ L., Papovich, C., Ryan, R.\ E., et al.\ 2012b, \apj, 758, 93
\bibitem[Giavalisco(2002)]{giavalisco02} Giavalisco, M. 2002, \araa, 40, 579
\bibitem[Giavalisco et al.(2004)]{giavalisco04} 
	Giavalisco, M., Ferguson, H.\ C., \& Koekemoer, A.\ M.\, et al.\ 2004, \apj, 600, L93
\bibitem[Gonzales-Perez et al.(2013)]{gonzales-perez13}
	Gonzales-Perez, V., Lacey, C.\ G., Baugh, C.\ M., Frenk, C.\ S., \& Wilkins, S.\ M.\ 2013, \mnras, 429, 1609
\bibitem[Greve(2010)]{greve10}
	Greve, T. R., et al. 2010, \apj, 719, 483
\bibitem[Hathi, Malhotra, \& Rhoads(2008)]{hathi08}
	Hathi, N.\ P., Malhotra, S., \& Rhoads, J.\ E.\ 2008, \apj, 673, 686
\bibitem[Hathi et al.(2012)]{hathi12}
	Hathi, N.\ P., Mobasher, B., Capak, P., Wang, W.-H., \& Ferguson, H.\ C.\ 2012, \apj, 757, 43
\bibitem[Helou, Soifer, \& Rowan-Robinson(1985)]{helou85}
   	Helou, G., Soifer, B.\ T., \& Rowan-Robinson, M.\ 1985, \apjl, 298, L7
\bibitem[Ho et al.(2010)]{ho10}
   	Ho, I.-T., Wang, W.-H., Morrison, G.\ E., \& Miller, N.\ A.\ 2010, \apj, 722, 1051 (Ho10)
\bibitem[Hsieh et al.(2012)]{hsieh12}
	Hsieh, B.-C., Wang, W.-H., Yan, H., et al.\ 2012, \apj, 749, 88
\bibitem[Kurczynski \& Gawiser(2010)]{kurczynski10}
	Kurczynski, P., Gawiser, E. 2010, \aj, 139, 1592
\bibitem[Kennicutt(1998)]{Kennicutt98}
	Kennicutt, R.\ C.\ 1998, \araa, 36, 189
\bibitem[Kennicutt \& Evans(2012)]{Kennicutt12}
	Kennicutt, R.\ C.\ \& Evans, N.\ J.\ 2012, \araa, 50, 531
\bibitem[Lee et al.(2012)]{lee12}
   	Lee, K.-S, Alberts, S., Atlee, D., et al.\ 2012, \apjl, 758, L31
\bibitem[Lehmer et al.(2005)]{lehmer05}
	Lehmer, B.\ D., Brandt, W.\ N., \& Alexander, D.\ M.\ 2005, \aj, 129, 1
\bibitem[Madau(1995)]{madau95} Madau, P.\ 1995, \apj, 441, 18
\bibitem[Madau et al.(1998)]{Madau98}
	Madau, P.,Pozzetti, L., \& Dickinson, M. 1998, \apj , 498, 106
\bibitem[McLure et al.(2010)]{mclure10}
	McLure, R.\ J., Dunlop, J.\ S., Cirasuolo, M.\ 2010, \mnras, 403, 960
\bibitem[Meurer et al.(1999)]{meurer99}
	Meurer, G.\ R., Heckman, T. M., \& Calzetti, D. 1999, \apj , 521, 64 (M99)
\bibitem[Miller et al.(2008)]{miller08}
	Miller, N.\ A., Fomalont, E.\ B., Kellermann, et al.\ 2008, \apjs, 179, 114
\bibitem[Morrison et al.(2010)]{morrison10}
   	Morrison, G.\ E., Owen, F.\ N., Dickinson, M., Ivison, R.\ J., \& Ibar, E.\ 2010, \apjs, 188, 178
\bibitem[Oesch et al.(2010)]{oesch10}
   	Oesch, P.\ A., et al.\ 2010, \apjl, 709, L16
\bibitem[Oesch et al.(2013)]{oesch13}
   	Oesch, P.\ A., et al.\ 2013, \apj, 772, 136
\bibitem[Ono et al.(2012)]{ono12}
	Ono, Y., Ouchi, M., \& Mobasher, B., et al.\ 2012, \apj, 744, 83
\bibitem[Oteo et al.(2013)]{oteo13}
	Oteo, I., Cepa, J., \& Bongiovanni, \'{A}., et al.\ 2013, \aap, 554, 3
\bibitem[Ouchi et al.(2004)]{ouchi04}
	Ouchi, M., Shimasaku, K., Okamura, S., et al.\ 2004, \apj, 611, 660
\bibitem[Ouchi et al.(2012)]{ouchi09}
	Ouchi, M., Mobasher, B., Shimasaku, K., et al.\ 2009, \apj, 706, 1136
\bibitem[Overzier et al.(2011)]{overzier11}
	Overzier, R.\ A., Heckman, T.\ M., Wang, J., et al.\ 2011, \apjl, 726, L7
\bibitem[Peacock et al.(2000)]{peacock00}
	Peacock, J.\ A., et al.\ 2000, \mnras, 318, 535
\bibitem[Pei(1992)]{pei92}
   	Pei, Y.\ C.\ 1992, \apj, 395, 130
\bibitem[Perera et al.(2008)]{perera08}
   	Perera, T.\ A., Chapin, E.\ L., Austermann, J.\ E., et al.\ 2008, \mnras, 391, 1227
\bibitem[Planck Collaboration(2013)]{planck13}
	Planck Collaboration 2013, \aap, submitted (arXiv:1303.5076)
\bibitem[Rau \& Cornwell(2011)]{rau2011}
   	Rau, U., \& Cornwell, T.\ J.\ 2011, \aap, 532, 71
\bibitem[Reddy, \& Steidel(2004)]{reddy04}
	Reddy, N.\ A., \& Steidel, C.\ C.\ 2004, \apj, 603, L13
\bibitem[Reddy, \& Steidel(2006)]{reddy06}
	Reddy, N.\ A., Steidel, C.\ C., \& Fadda, D., et al.\ 2006, \apj, 644, 792
\bibitem[Reddy et al.(2010)]{reddy10}
   	Reddy, N.\ A., Erb, D.\ K., Pettini, M., Steidel, C.\ C., Shapley, A.\ E.\ 2010, \apj, 712, 1070
\bibitem[Reddy et al.(2012a)]{reddy12a}
   	Reddy, N., Dickinson, M., Elbaz, D., et al.\ 2012a, \apj, 744, 154
\bibitem[Reddy et al.(2012b)]{reddy12b}
   	Reddy, N.\ A., Pettini, M., Steidel, C. C., et al.\ 2012b, \apj, 754, 25
\bibitem[Rigopoulou et al.(2010)]{rigopoulou10}
	Rigopoulou, D., Magdis, G., Ivison, R.\ J., et al.\ 2010, \mnras, 409, L7
\bibitem[Robertson et al.(2013)]{robertson13}
	Robertson, B.\ E., Furlanetto, S.\ R., Schneider, E., et al.\ 2013, \apj, 768, 71
\bibitem[Schenker et al.(2012)]{schenker12}
	Schenker, M.\ A., Stark, D.\ P., Ellis, R.\ S., et al.\ 2012, \apj, 744, 179
\bibitem[Sklias et al.(2014)]{sklias14}
	Sklias, P., Zamojski, M., Schaerer, D., et al.\ 2014, \aap, 561, A149
\bibitem[Songaila et al.(1990)]{songaila90} 
	Songaila, A., Cowie, L.\ L. \& Lilly, S.\ J. 1990, \apj, 348, 371
\bibitem[Stanway, Bunker, \& McMahon(2003)]{stanway03}
 	Stanway, E.\ R., Bunker, A.\ J., \& McMahon, R.\ G.\ 2003, \mnras, 342, 439
\bibitem[Stanway, McMahon, \& Bunker(2005)]{stanway05}
 	Stanway, E.\ R., McMahon, R.\ G., \& Bunker, A.\ J.\ 2005, \mnras, 359, 1184
\bibitem[Steidel \& Hamilton(1992)]{steidel92} 
	Steidel, C.\ C. \& Hamilton, D.\ 1992, \aj, 104, 941
\bibitem[Steidel \& Hamilton(1993)]{steidel93} 
	Steidel, C.\ C. \& Hamilton, D.\ 1993, \aj, 105, 2017 
\bibitem[Steidel et al.(1995)]{steidel95} 
	Steidel, C.\ C., Pettini, M. \& Hamilton, D.\ 1995, \aj, 110, 2519
\bibitem[Steidel et al.(1999)]{steidel99}
 	Steidel, C.\ C., Adelberger, K.\ L. Giavalisco, M., Dickinson, M., \& Pettini, M.\ 1999, \apj, 519, 1
\bibitem[Steidel et al.(2003)]{steidel03}
	Steidel, C.\ C., Adelberger, K.\ L., Shapley, A.\ E., et al.\ 2003, \apj, 592, 728
\bibitem[Takeuchi et al.(2010)]{takeuchi10}
	Takeuchi, T.\ T., Buat, V., Heinis, S., et al.\ 2010, \aap, 514, A4
\bibitem[Takeuchi et al.(2012)]{takeuchi12}
	Takeuchi, T.\ T., Yuan, F.-T., Ikeyama, A., Murata, K.\ L., \& Inoue, A.\ K.\ 2012, \apj, 755, 144
\bibitem[Treu et al.(2013)]{treu13}
	Treu, T., Schmidt, K.\ B., Trenti, M., Bradley, L.\ D., \& Stiavelli, M.\ 2013, \apj, 775, L29
\bibitem[Vijh, Witt, \& Gordon(2003)]{vijh03}
   	Vijh, U.\ P., Witt, A.\ N., \& Gordon, K.\ D.\ 2003, \apj, 587, 533
\bibitem[Verma et al.(2007)]{verma07}
   	Verma, A., Lehnert, M.\ D., F\"{o}rster Schreiber, N.\ M., Bremer, M.\ N., \& Douglas, L.\ 2007, \mnras, 377, 1024
\bibitem[Wang \& Heckman(1996)]{wang96}
	Wang, B.\, \& Heckman, T.\ M.\ 1996, \apj, 457, 645
\bibitem[Wang, Barger, \& Cowie(2009)]{wang09}
   	Wang, W.-H., Barger, A.\ J., \& Cowie, L.\ L.\ 2009, \apj, 690, 319
\bibitem[Wang, Barger, \& Cowie(2012)]{wang12}
   	Wang, W.-H., Barger, A.\ J., \& Cowie, L.\ L.\ 2012, \apj, 744, 155
\bibitem[Webb et al.(2003)]{webb03}
   	Webb T.\ M.\ A., Eales, S., \& Lilly S.\ J.\ 2003, \apj, 582, 6
\bibitem[Webb et al.(2004)]{Webb04}
   	Webb T.\ M.\ A., Brodwin M., Eales S., \& Lilly S.\ J.\ 2004, \apj, 605, 645
\bibitem[Wilkins et al.(2013)]{wilkins13}
   	Wilkins, S.\ M., Bunker, A., Coulton, W., et al.\ 2013, \mnras, 430, 2885
\bibitem[Yan \& Windhorst(2004)]{yan04}
	Yan, H., \& Windhorst, R.\ A.\ 2004, \apj, 612, L93
\end{thebibliography}
\end{document}